\documentclass[twocolumn,pra,floatfix,citeautoscript,nofootinbib,superscriptaddress]{revtex4}
\usepackage{amsbsy}
\usepackage{latexsym,epsfig,graphicx}
\usepackage{dcolumn}
\usepackage{graphicx}
\usepackage{subfigure}
\usepackage{comment}
\usepackage{color}
\usepackage{bm}
\usepackage{mathrsfs}
\usepackage{amsfonts}
\usepackage{amsmath}
\usepackage{color}
\usepackage{amssymb}
\usepackage{xspace}
\usepackage{epstopdf}
\usepackage{tabularx}
\usepackage{longtable}
\usepackage[colorlinks=true, letterpaper=true, pdfstartview=FitV, linkcolor=blue, citecolor=blue, urlcolor=blue]{hyperref}
\usepackage[normalem]{ulem}

\pdfoutput=1
\input{tcilatex}
\begin{document}

\title{Wannier-type photonic higher-order topological corner states induced
solely by gain and loss}
\author{Ya-Jie Wu}
\thanks{wuyajie@xatu.edu.cn}
\affiliation{School of Science, Xi'an Technological University, Xi'an 710032, China}
\affiliation{Department of Physics, The University of Texas at Dallas, Richardson, Texas
75080-3021, USA}
\author{Chao-Chen Liu}
\affiliation{School of Science, Xi'an Technological University, Xi'an 710032, China}
\author{Junpeng Hou}
\thanks{junpeng.hou@utdallas.edu}
\affiliation{Department of Physics, The University of Texas at Dallas, Richardson, Texas
75080-3021, USA}

\begin{abstract}
Photonic crystals have provided a controllable platform to examine
excitingly new topological states in open systems. In this work, we reveal
photonic topological corner states in a photonic graphene with
mirror-symmetrically patterned gain and loss. Such a nontrivial Wannier-type
higher-order topological phase is achieved through solely tuning on-site
gain/loss strengths, which leads to annihilation of the two valley Dirac
cones at a time-reversal-symmetric point, as the gain and loss change the
effective tunneling between adjacent sites. We find that the
symmetry-protected photonic corner modes exhibit purely imaginary energies
and the role of the Wannier center as the topological invariant is illustrated. For
experimental considerations, we also examine the topological interface
states near a domain wall. Our work introduces an interesting platform for
non-Hermiticity-induced photonic higher-order topological insulators,
which, with current experimental technologies, can be readily accessed.
\end{abstract}

\maketitle

\section{Introduction}

Topological phases and topological phase transitions in fermionic and
bosonic systems, described by Hermitian Hamiltonians, have attracted great
interest in the past three decades \cite{Hasan2010,Qi2011,Ozawa2019}. Recent
studies have revealed that topological phases can be extended to
non-Hermitian systems beyond the scope of closed systems \cite%
{Rudner2009,Regensburger2013,Lee2016,Leykam2017,XuY2017,JinL2017,Shen2018,Flore2018,Yao2018a,Yao2018b,Gong2018,Yokomizo2019,Turker2019,JinL2019,LiuC2019}%
. Especially, the interplay between non-Hermiticity and topological states
leads to unique properties that have no counterparts in Hermitian systems
\cite{Kawabata2019,Lieu2018}. While the non-Hermitian parameters are hard to
tune in most classical or quantum systems, optical and photonic systems \cite%
{Poli2015,Weimann2017,Feng2017,Xiao2017,Miri2019,St-Jean2017,Zhao2018}
provide controllable platforms to investigate non-Hermitian physics, in
which the real and imaginary parts of the eigenenergy of a photonic mode are
related to its frequency and amplifications/attenuations over time.

Photonic graphene, described by a two-dimensional (2D) honeycomb lattice
consisting of optical cavities, exhibits various interesting features and
thus has been intensively studied in past decades \cite%
{Efremidis2002,Bartal2005,Peleg2007,Sepkhanov2007,Treidel2008,Treidel2010,Polini2013,Plotnik2014,Oztas2018,Ozawa2019}%
. It also enjoys significant advantages in terms of tunability of the lattice
geometry and cleanness (absence of disorder or nonlinear interaction).
Photonic graphene mimics a semimetal with two inequivalent gapless Dirac
cones carrying opposite Berry phases $\pm \pi $. The local stability of
Dirac points is guaranteed by time-reversal as well as inversion symmetries,
while the global stability is protected by $C_{3}$ symmetry. By projecting
Dirac points onto zigzag edges, the chiral surface states are raised, which
have been observed in both electronic and photonic systems \cite%
{Kobayashi2005,Rechtsman2013}. In particular, the artificial photonic
lattices provide an ideal platform for the simulation of non-equilibrium
open systems with gain and loss. Photonic graphene with balanced gain
and loss has been studied in previous work, and it is found that
topologically protected edge states with non-degenerate purely imaginary
energies appear along the zigzag edges \cite{Oztas2018}.

More recently, a new type of topological phase, dubbed a higher-order
topological state, was proposed \cite%
{Benalcazar2017,Langbehn2017,Song2017,Ezawa2018}. Photonic systems provide a
powerful experimental platform for higher-order topological insulators \cite%
{Noh2018,ChenX2019,Hassan2019,Xie2018}. Formally, $d$-dimensional, $r^{%
\mathrm{th}}$-order topological phases host ($d-r$)D topologically-protected
edge states. For instance, a 2D/3D second-order topological insulator (SOTI)
hosts zero-energy corner/hinge states while the $r=1$ cases reduce to the
conventional topological insulators. The fate of higher-order topological
corner states in open systems then becomes an important and intriguing
question. Previous studies have employed either asymmetric intracell hopping
\cite{liut2019,Lee2019} or onsite gain/loss \cite{Luo2019} to induce corner
states based upon the 2D generalization of Su-Schrieffer-Heeger model on a
square lattice \cite{Benalcazar2017}.

In this paper, we enrich the family of non-Hermitian photonic SOTIs by
proposing a minimal Wannier-type SOTI, solely induced by mirror-symmetric
gain and loss, in photonic graphenes. We show that the photonic semimetal
phase can be driven into a Wannier-type SOTI phase through tuning a stronger
onsite gain/loss rate. Such a system hosts photonic corner modes and is
characterized by a nontrivial topological invariant, known as the Wannier
center or photonic polarization.

The paper is organized as follows. In Sec. \ref{Sec2}, we introduce a
photonic graphene lattice with mirror-symmetric gain/loss and analyze
relevant symmetries of the non-Hermitian Hamiltonian. In Sec. \ref{Sec3}, we
study how the band structure changes with respect to increasing gain/loss
strength and show there is a topological phase transition from a photonic
semimetal to a photonic Wannier-type SOTI, characterized by a nontrivially
quantized Wannier center \cite{Ezawa2018,Benalcazar2019}. In Sec. \ref{Sec4}%
, we examine the photonic higher-order topological corner states, and
topological interface states at the domain wall of two non-Hermitian
photonic graphene. Finally, we make relevant discussions and draw
conclusions in Sec. \ref{Sec5}.

\section{Photonic graphene with a patterned gain and loss}

\label{Sec2}

\begin{figure}[tbp]
\centering\includegraphics[width=0.42\textwidth]{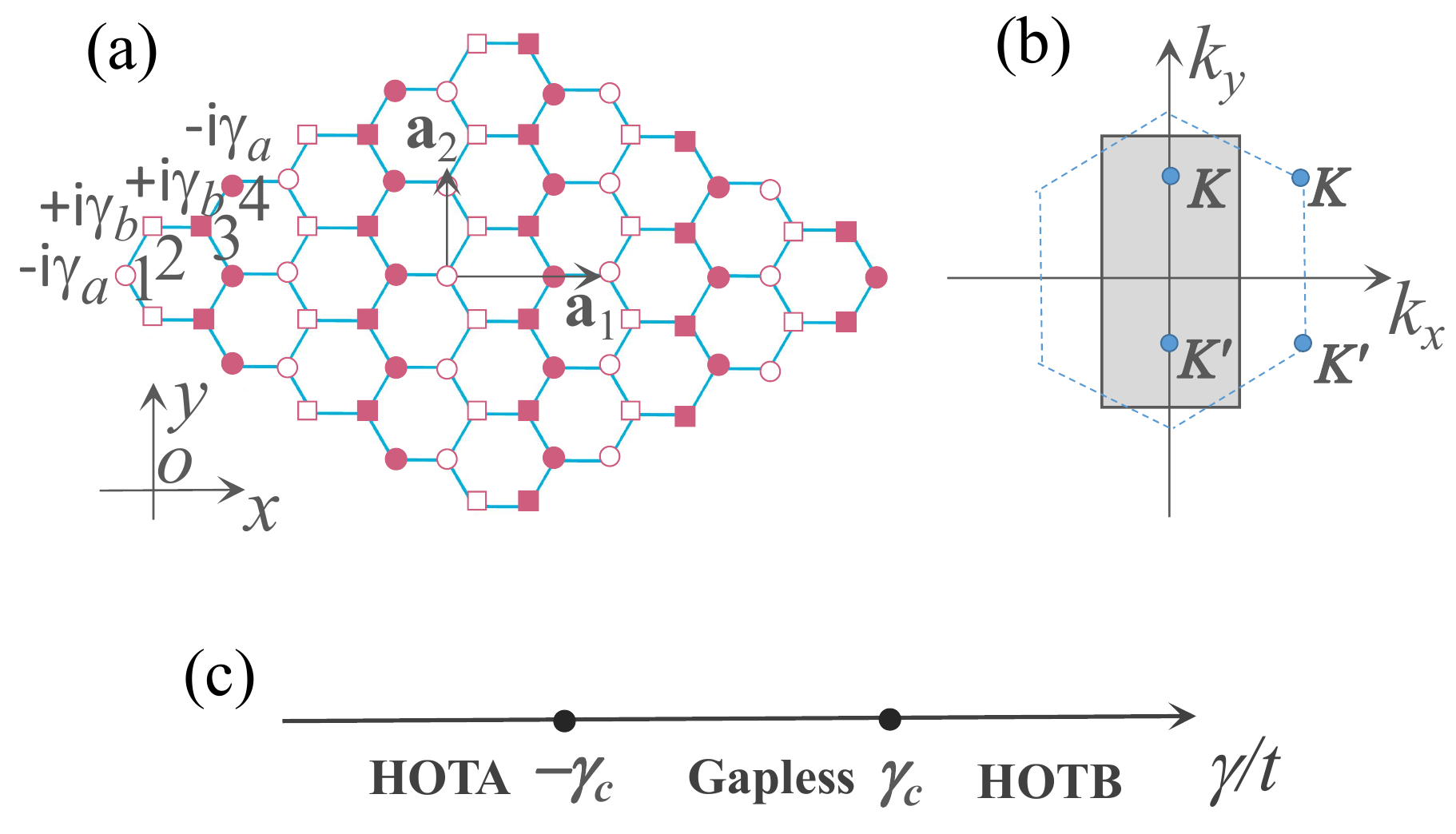}
\caption{(a) Illustration of photonic graphene lattice with patterned
gain/loss $+i\protect\gamma _{b}$ and $-i\protect\gamma _{a}$. One unitcell
consists of four sublattices indexed by $1$ (circle), $2$ (hollow square), $%
3 $ (solid square), and $4$ (disk). (b) The dashed green curves sketch the
first Brillouin zone of the original honeycomb lattice without gain and loss,
while the gray shaded area indicates the first Brillouin zone corresponding
to the enlarged unit cells of gain/loss-dressed lattice. (c) The phase
diagram vs the strength of gain and loss $\protect\gamma $ in units of
the bare nearest-neighbor hopping $t$.}
\label{Fig1}
\end{figure}

We consider uniformly coupled optical cavities on a honeycomb lattice,
forming the photonic graphene. Each unitcell consists of four coupled
resonators as shown in Fig. \ref{Fig1} (a). A mirror-symmetric imaginary
onsite potential $(-i\gamma _{a},i\gamma _{b},i\gamma _{b},-i\gamma _{a})$
is introduced along $x$ and we denote $t$ as the bare coupling between
nearest-neighbor cavities $\left\langle i_{m},j_{n}\right\rangle $. The
single-particle Hamiltonian is written as $H=H_{0}+H_{\mathrm{NH}}$ and the
Hermitian part reads $H_{0}=-t\sum_{\left\langle i_{m},j_{n}\right\rangle
}\left( a_{i_{m}}^{\dagger }a_{j_{n}}+h.c.\right) $. The non-Hermitian
potential is $H_{\mathrm{NH}}=-i\sum_{m,i}\gamma _{m}a_{i_{m}}^{\dagger
}a_{i_{m}}$, where $\gamma _{m=1,4}=\gamma _{a}$ and $\gamma
_{m=2,3}=-\gamma _{b}$ describe the patterned gain and loss. The
operator $a_{i_{m}}^{\dagger }$ creates a photonic cavity mode at site $%
i_{m} $. Note that the non-Hermitian potential enlarges the unitcell of the
usual honeycomb lattice, and the Bravais vectors are now given by $\mathbf{a}%
_{1}=\left( 3,0\right) $ and $\mathbf{a}_{2}=\left( 0,\sqrt{3}\right) $, as
illustrated in Fig. \ref{Fig1}(a). The first Brillouin zone decreases
correspondingly and it is sketched in Fig. \ref{Fig1}(b). Here we set the
lattice spacing of the graphene lattice to be a unit. The total Hamiltonian $H$
in momentum space can be written as $H=\sum_{k}\Psi _{k}^{\dagger }H\left(
k\right) \Psi _{k}$ in the basis $\Psi _{k}=\left(
a_{1,k},a_{3,k},a_{2,k},a_{4,k}\right) ^{T}$ and
\begin{eqnarray}
H\left( k\right) &=&\func{Re}\left( \xi _{k}\right) \sigma _{x}\tau _{0}-%
\func{Im}\left( \xi _{k}\right) \sigma _{y}\tau _{0}+\func{Re}\left( \beta
_{k}\right) \sigma _{x}\tau _{x}  \notag \\
&&-\func{Im}\left( \beta _{k}\right) \sigma _{y}\tau _{x}-i\gamma _{+}\sigma
_{z}\tau _{z}-i\gamma _{-}\sigma _{0}\tau _{0},
\end{eqnarray}%
where $\func{Re}\left( \xi _{k}\right) =-2t\cos \left( \sqrt{3}%
k_{y}/2\right) \cos \left( k_{x}/2\right) $, $\func{Im}\left( \xi
_{k}\right) =-2t\cos \left( \sqrt{3}k_{y}/2\right) \sin \left(
k_{x}/2\right) $, $\func{Re}\left( \beta _{k}\right) =-t\cos k_{x}$, $\func{%
Im}\left( \beta _{k}\right) =-t\sin k_{x}$ and $\gamma _{\pm }=\left( \gamma
_{a}\pm \gamma _{b}\right) /2$. The notations $\sigma _{x,y,z}$ and $\tau
_{x,y,z}$ represent Pauli matrices while $\sigma _{0}$ and $\tau _{0}$ are
identity matrices.

In the absence of gain and loss ($\gamma _{\pm }=0$), the Hamiltonian $H$
describes the well-known graphene model with two stable Dirac points
carrying geometric phases $\pm \pi $. The local stability of the Dirac
points is protected by inversion symmetry ($\mathcal{P}$) and time-reversal
symmetry ($\mathcal{T}$) while the global stability is guaranteed by $C_{3}$
rotational symmetry. In the presence of gain and loss, the $C_{3}$ symmetry
is broken, but both $\mathcal{P}$ and $\mathcal{T}$ symmetries are respected
when $\gamma _{a}=\gamma _{b}$. Hence, the system preserves $\mathcal{PT}$
symmetry as $(\mathcal{PT})H\left( k\right) \left( \mathcal{PT}\right)
^{-1}=H\left( k\right) $ with $\mathcal{PT=}\sigma _{x}\tau _{0}\mathcal{K}$%
, where $\mathcal{K}$ is complex-conjugation operator. In addition, there
exist mirror symmetries along $x$ and $y$, $\mathcal{M}_{x,y}H\left(
k_{x,y}\right) \mathcal{M}_{x,y}^{-1}=H\left( -k_{x,y}\right) $ with $%
\mathcal{M}_{x}=\sigma _{x}\tau _{x}$ and $\mathcal{M}_{y}=\sigma _{0}\tau
_{0}$, where the former is crucial for the topological characterization of
the nontrivial HOTI phase.

Without loss of generality, we consider the special case $\gamma _{a}=\gamma
_{b}=\gamma $, namely, $\gamma _{-}=0$ and $\gamma _{+}=\gamma $. The four
bands in momentum space can be analytically solved as
\begin{eqnarray}
E_{\pm ,+}\left( k\right) &=&\pm \sqrt{3t^{2}+2t^{2}\cos \left( \sqrt{3}%
k_{y}\right) -\gamma ^{2}+2\sqrt{e_{k}}},  \notag \\
E_{\pm ,-}\left( k\right) &=&\pm \sqrt{3t^{2}+2t^{2}\cos \left( \sqrt{3}%
k_{y}\right) -\gamma ^{2}-2\sqrt{e_{k}}},  \label{Eq1}
\end{eqnarray}%
where $e_{k}=4t^{4}\cos ^{2}\left( \sqrt{3}k_{y}/2\right) \cos ^{2}\left(
3k_{x}/2\right) -t^{2}\gamma ^{2}$. Compared to the two bands in a usual
graphene model, the energy bands here are folded and the Dirac points shift
to $\mathbf{K}$ and $\mathbf{K}^{\prime }$ due to the enlarged unit cell, as
shown in Figs. \ref{Fig1}(b) and \ref{Fig2}(a$_{1}$). In the presence of
gain and loss, there exist energy degeneracies $E_{+,+}=E_{+,-}$ and $%
E_{-,+}=E_{-,-}$ whenever $e_{k}=0$. This corresponds to an exceptional ring
in momentum space, which is given by $k_{\mathrm{E}}=2\arccos \left[ \gamma
/\left( 2t\cos \left( 3k_{x}/2\right) \right) \right] /\sqrt{3}$ and is
depicted in Figs.~\ref{Fig2}(b$_{2}$) and (c$_{2}$). For the eigenmodes
inside the exceptional ring, we have $e_{k}>0$ while $e_{k}<0$ otherwise.

\section{Topological phase transitions and topological invariants}

\label{Sec3}
\begin{figure}[tbp]
\centering\includegraphics[width=0.48\textwidth]{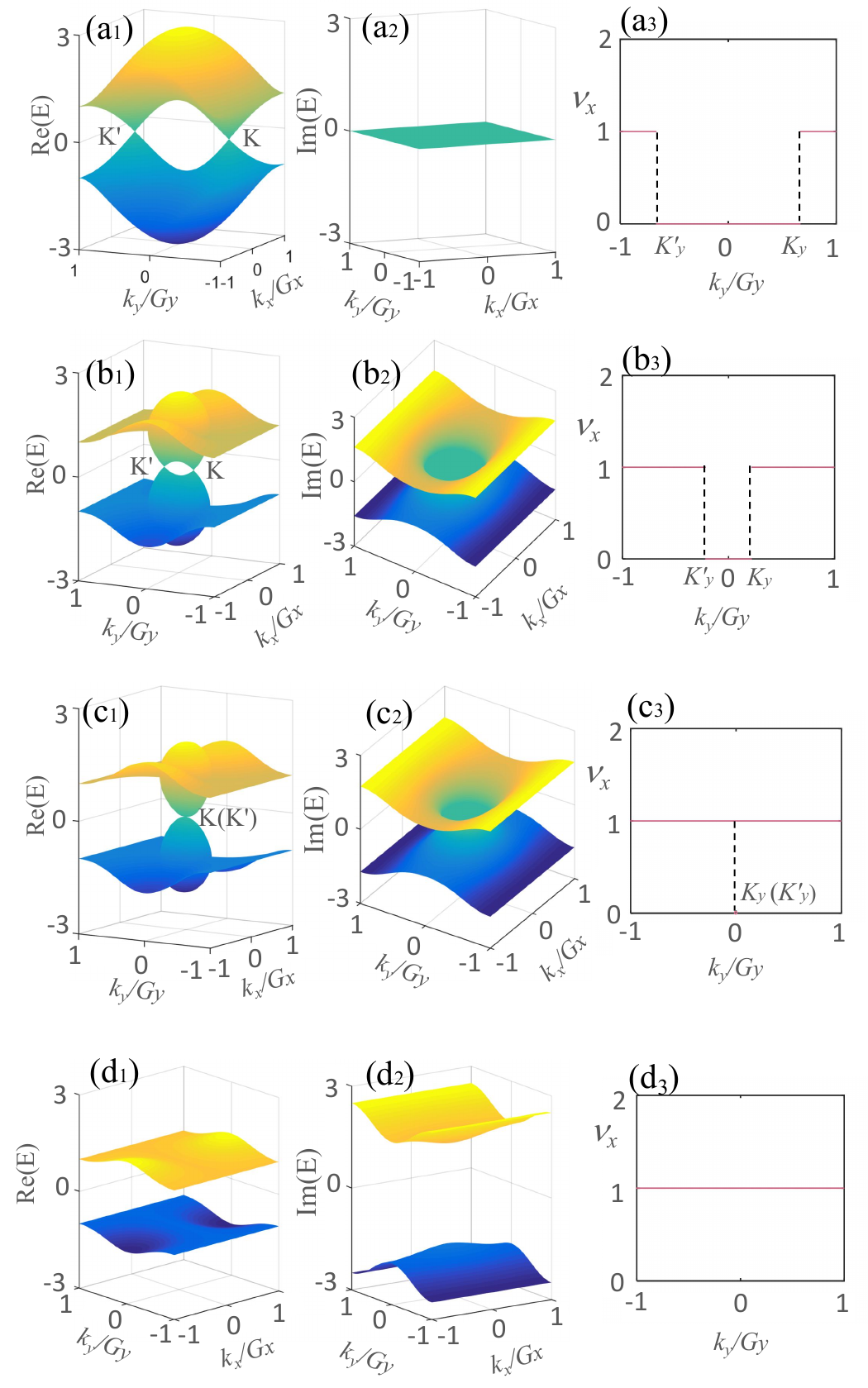}
\caption{Real spectrum (first column), imaginary spectrum (second column),
and topological invariant (third column) at different gain/loss strengths.
The non-Hermitian strength for each case is (a$_{\text{1}}$-a$_{\text{3}}$%
) $\protect\gamma =0$, (b$_{\text{1}}$-b$_{\text{3}}$) $\protect\gamma =1.6$%
, (c$_{\text{1}}$-c$_{\text{3}}$) $\protect\gamma =1.732$, and (d$_{\text{1}}$%
-d$_{\text{3}}$) $\protect\gamma =2.5$. Other parameters are chosen as $t=1$%
, $G_{x}=\protect\pi /3$, and $G_{y}=\protect\pi /\protect\sqrt{3}$.}
\label{Fig2}
\end{figure}

For each energy band $E_{\pm ,\pm }\left( k\right) $, we could
rewrite it as $E_{\pm ,\pm }\left( k\right) =\func{Re}\left[ E_{\pm ,\pm
}\left( k\right) \right] +i\func{Im}\left[ E_{\pm ,\pm }\left( k\right) %
\right] $. When the graphene lattice is free of gain and loss ($\gamma =0$)
, the energy bands are purely real and exhibit two Dirac points at $\mathbf{K%
}=\left( 0,2\sqrt{3}\pi /9\right) $ and $\mathbf{K}^{\prime }=\left( 0,-2%
\sqrt{3}\pi /9\right) $ [see Figs. \ref{Fig2}(a$_{\text{1}}$) and (a$_{\text{%
2}}$)]. As $\gamma $ increases, the exceptional ring gradually decreases,
and the two Dirac points remain massless yet their locations in momentum
space change. More explicitly, with larger $\gamma $, two Dirac points
approach each other gradually [see Figs. \ref{Fig2}(b$_{\text{1}}$) and (b$_{%
\text{2}}$)] and then merge at the time-reversal-invariant point $\Gamma
=\left( 0,0\right) $ at critical gain/loss strength $\gamma _{c}=1.732$ [see
Figs. \ref{Fig2}(c$_{\text{1}}$) and (c$_{\text{2}}$)]. Finally, when $\gamma
>\gamma _{c}$ there opens a real energy gap, which is plotted in Figs.~\ref%
{Fig2}(d$_{\text{1}}$) and (d$_{\text{2}}$). Later, we will show that the
phase transition described above has a topological nature.

Based upon the symmetry analyses in Sec. \ref{Sec2}, the Hamiltonian $H(k)$
preserves both $\mathcal{PT}$ symmetry and mirror symmetry along $x$. In
particular, the reflection symmetry $M_{x}$ leads to the nontrivial
quantization of \textquotedblleft photonic polarization\textquotedblright\
along the $x$ direction. To formulate this as a bulk property, we construct a
Wilson loop operator in $x$ direction $W_{x,k}$, where $k$ denotes the
starting (base) point of the loop. We denote the Bloch wave function of the
occupied energy bands with negative real energies as $\left\vert
u_{m,k}^{R,L}\right\rangle $, where $\left\vert u_{m,k}^{L}\right\rangle $
and $\left\vert u_{m,k}^{R}\right\rangle $ are left and right eigenvectors
defined as $H^{\dagger }\left( k\right) \left\vert u_{m,k}^{L}\right\rangle
=E_{m}^{\ast }\left( k\right) \left\vert u_{m,k}^{L}\right\rangle $ and $%
H\left( k\right) \left\vert u_{m,k}^{R}\right\rangle =E_{m}\left( k\right)
\left\vert u_{m,k}^{R}\right\rangle $ with normalization condition $%
\left\langle u_{m,k}^{L}|u_{n,k^{\prime }}^{R}\right\rangle =\delta
_{m,n}\delta _{k,k^{\prime }}$. We define $\left[ F_{x,k}\right]
^{m,n}=\left( \left\langle u_{m,k+\Delta k_{x}}^{L}|u_{n,k}^{R}\right\rangle
+\left\langle u_{m,k+\Delta k_{x}}^{R}|u_{n,k}^{L}\right\rangle \right) /2$,
where $\Delta k_{x}=2\pi /N_{x}$ with $N_{x}$ the number of lattice sites
along the $x$ direction \cite{Hou2019}. The Wilson loop operator is then $%
W_{x,k}=F_{x,k+N_{x}\Delta k_{x}}...F_{x,k+\Delta k_{x}}F_{x,k}$. We define
the topological invariant $v_{x}\left( k_{y}\right) =-\frac{i}{\pi }\mathrm{%
Tr}\left( \ln W_{x,k}\right) $, forming the Wannier band. It is quantized
under reflection symmetries. In the thermodynamic limit, the topological
invariant at each $k_{y}$ is%
\begin{equation}
v_{x}\left( k_{y}\right) =-\frac{1}{\pi }\mathrm{Tr}\left( \doint \mathcal{A}%
_{k}dk_{x}\right) ,
\end{equation}%
where $\left( \mathcal{A}_{k}\right) _{m,n}=\left( \mathcal{A}_{k}^{LR}+%
\mathcal{A}_{k}^{RL}\right) /2$ is non-Abelian Berry connection with $\left(
\mathcal{A}_{k}^{\alpha \beta }\right) _{mn}=-i\left\langle u_{m,k}^{\alpha
}|\partial _{k_{x}}u_{n,k}^{\beta }\right\rangle $ and $\alpha ,\beta =L,R$.
Following similar steps, we could obtain the topological invariant $%
v_{y}\left( k_{x}\right) $ at each $k_{x}$. Finally, the topological
invariant (Wannier center) of Wannier bands is defined as $\left(
v_{x}^{\prime },v_{y}^{\prime }\right) $ with $v_{x/y}^{\prime }=\frac{1}{%
4G_{y/x}}\doint v_{x/y}\left( k_{y/x}\right) dk_{y/x}$, where $G_{x}=\pi /3$
and $G_{y}=\pi /\sqrt{3}.$

In the absence of gain and loss, topological invariants are $v_{x}\left(
k_{y}\right) =1$ if $-\pi /\sqrt{3}<k_{y}<-2\sqrt{3}\pi /9$ or $2\sqrt{3}\pi
/9<k_{y}<\pi /\sqrt{3}$ and $v_{x}\left( k_{y}\right) =0$ otherwise [see
Fig.~\ref{Fig2}(a$_{\text{3}}$)]. As the strength of gain and loss
increases, we see the region with $v_{x}\left( k_{y}\right) =1$ enlarges
along $k_{y}$ [see Fig.~\ref{Fig2}(b$_{\text{3}}$)]. At the critical point $%
\gamma _{c}$, $v_{x}\left( k_{y}\right) =1$ at any $k_{y}$ except the Dirac
points $\mathbf{K}$ ($\mathbf{K}^{\prime }$). Finally, when $\gamma >\gamma
_{c}$, the real-energy gap opens, as shown in Fig. ~\ref{Fig2}(d$_{\text{1}}$%
), with the topological invariant $v_{x}\left( k_{y}\right) =1$ for all $%
k_{y}$. At this time, the Wannier center of the Wannier bands is quantized
to a nontrivial value $\left( 1/2,0\right) $. Similar cases happen when $%
\gamma <$ $0$, and if $\gamma <-\gamma _{c}$, the Wannier center becomes $%
\left( -1/2,0\right) $.

We would like to remark that the quantization of topological invariant $%
v_{x}^{\prime }$ is guaranteed by mirror symmetry along the $x$ direction,
and is robust against weak mirror-symmetric perturbations (see Appendix \ref%
{app} for more details).

\section{Wannier-type topological corner modes and interface modes}

\label{Sec4}

\begin{figure}[h]
\centering\includegraphics[width=0.48\textwidth]{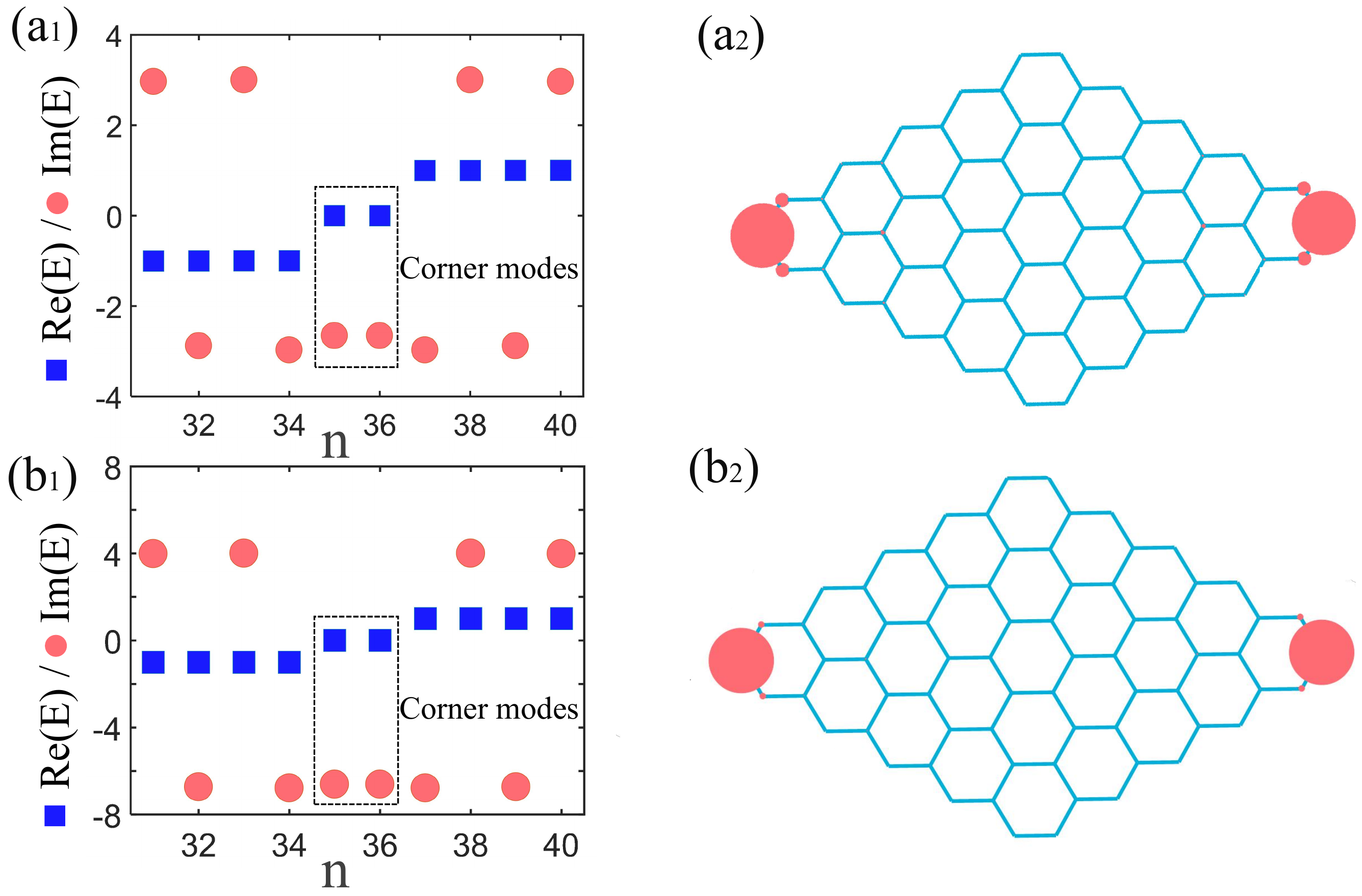}
\caption{(a$_{\text{1}}$) Real and imaginary parts of eigenspectrum [$\func{%
Re}\left( E\right)$ and $\func{Im}\left( E\right) $] vs the state index $%
n$. The two states in dashed rectangle are photonic corner modes with purely
imaginary energies. (a$_{\text{2}}$) Spatial density distribution of the
corner modes and the radii of the pink disk is proportional to local
density. We choose $t=1$ and $\protect\gamma _{a}=\protect\gamma _{b}=%
\protect\gamma =3$ in these two panels. (b$_{\text{1}}$, b$_{\text{2}}$%
) Similar to (a$_{\text{1}}$) and (a$_{\text{2}}$) but plotted with
different gain/loss profiles $\protect\gamma _{a}=4.0$ and $\protect\gamma %
_{b}=6.8$, demonstrating the robustness of the photonic topological corner
modes.}
\label{Corner}
\end{figure}

We focus on a sample shown in Fig. \ref{Fig1}(a) and tune the parameters so
that it is in topological phase with a Wannier center quantized to $\left(
1/2,0\right) $. We first consider the case $\gamma =3.0$ and the numeric
results are plotted in Fig. \ref{Corner}(a$_{\text{1}}$), where we observe
two degenerate modes with purely imaginary energies. The corresponding
particle density distributions are also plotted in Fig. \ref{Corner}(a$_{%
\text{2}}$), which shows that the two modes are localized at two horizonal
corners of the given sample. We refer to this phase as second-order
topological phase A (HOTA). Similarly, we find that the photonic topological
corner modes also emerge when $\gamma <-\gamma _{c}$. This phase is dubbed
second-order topological phase B (HOTB), and together with HOTA they are
illustrated in the phase diagram in Fig. \ref{Fig1}(c). So far, we have
focused on the special cases where $\gamma _{a}=\gamma _{b}$. However, from
previous discussions, we have argued that the quantization of a nontrivial
topological invariant is protected by the mirror symmetry, which is also
preserved when $\gamma _{a}\neq \gamma _{b}$. Thus, we would expect the
system to be nontrivial even when $\gamma _{a}\neq \gamma _{b}$, as long as
the bulk spectrum is gapped. We demonstrate this point by showing the
photonic corner modes in Figs. \ref{Corner}(b$_{\text{1}}$) and (b$_{\text{2}}
$), where we set $\gamma _{a}=4.0$ and $\gamma _{b}=6.8$. This suggests that
photonic corner modes are symmetry protected and they are robust against any
mirror-symmetric perturbation. Here, we also would like to remark that our
model also satisfies a pseudo-anti-Hermiticity, $H^{\dagger }\left( k\right)
=-\eta H\left( k\right) \eta $, where $\eta =\sigma _{z}\tau _{0}$. This
symmetry can lead to a nontrivial topology via chirality in terms of
pairwise eigenvalues, $E$ and $-E^{\ast }$. For a single corner state, it is
also the eigenstate of the operator $\eta $. Therefore, the eigenenergy of
the corner state satisfies $E=-E^{\ast }$, which suggests that the real
component of energy of the corner state must be pinned at zero.

\begin{figure}[tbp]
\centering\includegraphics[width=0.48\textwidth]{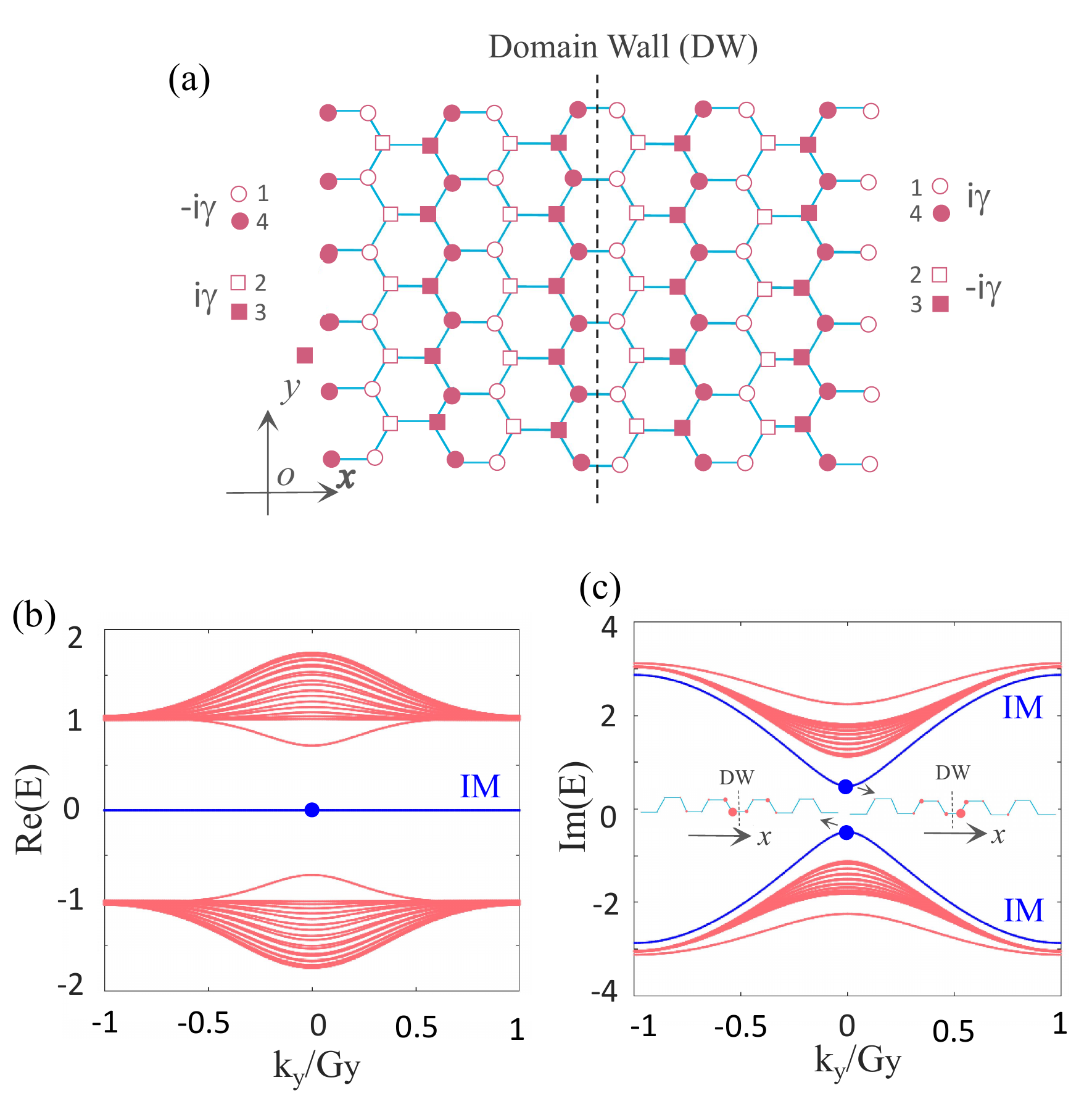}
\caption{(a) Two photonic graphene sheets with a domain wall (DW) in
between, which is highlighted by the dashed line. As they possess opposite
topological invariants, topological interface modes near the
domain wall are raised. (b, c) Real and imaginary spectrum along $k_{y}$, with an
open-boundary condition along $x$ (total $N_{x}=82$ sites), as specified in
panel (a). The blue curve indicates the interface mode (IM). The insets
depict the density distribution of two localized interface states.
Parameters are set to be $t=1$ and $\protect\gamma _{a}=\protect\gamma _{b}=%
\protect\gamma =3.2$.}
\label{domain}
\end{figure}

As discussed above, a photonic graphene with appropriate gain and loss $%
\gamma $ is a photonic higher-order topological insulator characterized by
topological invariant $\left( \text{sign}(\gamma )/2,0\right) $. In the
following, we consider two photonic graphene sheets separated by a domain
wall, which is depicted in Fig. \ref{domain}(a). While the translation
symmetry of graphene lattice is broken along $x$ direction, the translation
symmetry is respected along $y$. Hereafter, we consider $k_{y}$ as a system
parameter and treat $H(k_{y})$ as one dimension. The system Hamiltonian then
becomes
\begin{equation}
H=\sum_{i_{x}}H\left( i_{x}\right) =\sum_{i_{x}}H_{0}\left( i_{x}\right) +H_{%
\mathrm{NH}}\left( i_{x}\right) ,
\end{equation}%
with%
\begin{eqnarray}
H_{0}\left( i_{x}\right)
&=&-t\sum\nolimits_{k_{y}}a_{1,i_{x},k_{y}}^{\dagger
}a_{4,i_{x}-1,k_{y}}+\xi _{k_{y}}a_{1,i_{x},k_{y}}^{\dagger
}a_{2,i_{x},k_{y}}  \notag \\
&+&a_{2,i_{x},k_{y}}^{\dagger }a_{3,i_{x},k_{y}}+\xi _{k_{y}}^{\ast
}a_{3,i_{x},k_{y}}^{\dagger }a_{4,i_{x},k_{y}}+H.C., \\
H_{\mathrm{NH}}\left( i_{x}\right) &=&-i\sum_{m,k_{y}}\gamma _{m}\tilde{\eta}%
_{m}a_{m,i_{x},k_{y}}^{\dagger }a_{m,i_{x},k_{y}}
\end{eqnarray}%
where $\xi _{k_{y}}=t\left( 1+e^{i\sqrt{3}k_{y}}\right) $ and the domain
wall structure is given by $\tilde{\eta}_{m=1,4}=-\tilde{\eta}_{m=2,3}=1$
(left-hand side of the domina wall) and $\tilde{\eta}_{m=1,4}=-\tilde{\eta}%
_{m=2,3}=-1$ (right-hand side), as sketched in Fig. \ref{domain}(a).

The energy spectra of the system are given by $H\left( i_{x}\right)
\left\vert u\left( i_{x},k_{y}\right) \right\rangle $ $=E\left( k_{y}\right)
\left\vert u\left( i_{x},k_{y}\right) \right\rangle $. We set the strength
of gain and loss so that both photonic graphene sheets are in the
topological phases, where the left and right ones are HOTA and HOTB,
respectively. Through numeric calculations, we find localized states with
purely degenerate imaginary energy at the interface, i.e., the states with
negative (positive) imaginary energy localize at the left-hand (right-hand)
side of the domain wall, as shown in Fig. \ref{domain}(b) and (c). This
confirms that the two topological phases (HOTA and HOTB) indeed exhibit
different (in this case, opposite) topological properties and such a setup
can be used for experimental study of photonic HOTI in photonic graphene.

In such a domain-wall structure, we may also consider the general imaginary
potential $(-i\gamma _{a},i\gamma _{b},i\gamma _{b},-i\gamma _{a})$ with $%
\gamma _{a}\neq \gamma _{b}$ and the numeric calculations confirm the
existence of topological localized interface modes. Finally, we also
consider mirror-symmetric perturbations and find that, although the energies
of bulk states vary, the photonic topological interface states with purely
imaginary energy remain. In this sense, the topological interface states is
symmetry protected.

\section{Discussion and Conclusion}

\label{Sec5} Mirror-symmetric onsite gain/loss brings about the real-energy
gap in photonic graphenes and gives rise to localized photonic corner states
with purely imaginary energy. Physically, this arises from the fact that the
onsite imaginary potentials change the effective coupling between
nearest-neighbor cavities, rendering an effectively anisotropic $2$D
photonic crystals. With increasing strength of gain and loss, the effective
anisotropy of the photonic graphene grows. Correspondingly, the two Dirac
cones (with opposite Berry phases) approach each other and annihilate at a
high-symmetry point, leaving a gapped insulator phase, which tends out to be
a photonic Wannier-type HOTI phase \cite{Ezawa2018,Benalcazar2019}.

Our proposal provides a realistic scheme to realize photonic topological
corner states in photonic graphene. Moreover, it offers an accessible
platform to study higher-order generalization of the topological insulator
laser, which has been experimentally implemented in similar photonic crystal
\cite{Harari2018,Bandres2018,Parto2018,Bahari2017}. Topological corner modes
show negative imaginary parts. They can be promoted to a lasing mode using
high-Q cavities in visible or near-infrared range, and directly imaged in
real space when properly excited \cite{Bandres2018}.

We also remark that although the $\mathcal{PT}$ symmetry is respected in our
system when $\gamma _{a}=\gamma _{b}$, our discussion here is irrelevant to
the $\mathcal{PT}$-symmetric photonics studied in previous work \cite%
{Feng2017}. In their setups, while $\mathcal{PT}$ is respected, both $%
\mathcal{P}$ and $\mathcal{T}$ are broken individually. Our
gain/loss-dressed photonic graphene always preserves $\mathcal{P}$ (as well
as $\mathcal{M}_{x}$) symmetry even when $\gamma _{a}\neq \gamma _{b}$ and
it is known that spatial symmetry is crucial for stabilizing higher-order
corner modes \cite{Langbehn2017,Ezawa2018}. In previous works, the
topological properties of the $\mathcal{PT}$-symmetric systems usually
originate from the Hermitian parts of Hamiltonians\ \cite%
{Kunst2019,Zeuner2015,Yucea2018,Okugawa2019}, but in our work non-Hermitian
parts (patterned gain and loss) are crucial to induce the nontrivial phase.
Moreover, current $\mathcal{PT}$-symmetric topological systems concern
mainly $1$D systems and the discussion of $\mathcal{PT}$ symmetry is limited
to conventional topological phases \cite%
{Lee2016,Weimann2017,Parto2018,Kawabata2018,Okugawa2019}, but our model
presents a $2$D higher-order counterpart.

In summary, we have revealed Wannier-type higher-order topological states in
photonic graphenes with mirror-symmetric gain and loss. The
symmetry-protected topological corner modes and interface modes are robust
against perturbations respecting underlying symmetries. Our proposal does
not require finely tuned spacing or cavity structures and thus, can be
easily practiced in experiments.

\begin{acknowledgments}
This work is supported by the Scientific Research Program Funded by the Natural
Science Basic Research Plan in the Shaanxi Province of China (Programs No.
2018JQ1058 and No. 2019JM-001), the NSFC under the Grant No. 11504285, the
Scientific Research Program Funded by Shaanxi Provincial Education
Department under the grant No. 18JK0397, and the Young Talent fund of the
University Association for Science and Technology in Shaanxi, China (Program
No. 20170608).
\end{acknowledgments}

\appendix

\section{Robustness of corner states against perturbations}

\begin{figure}[tbp]
\centering\includegraphics[width=0.48\textwidth]{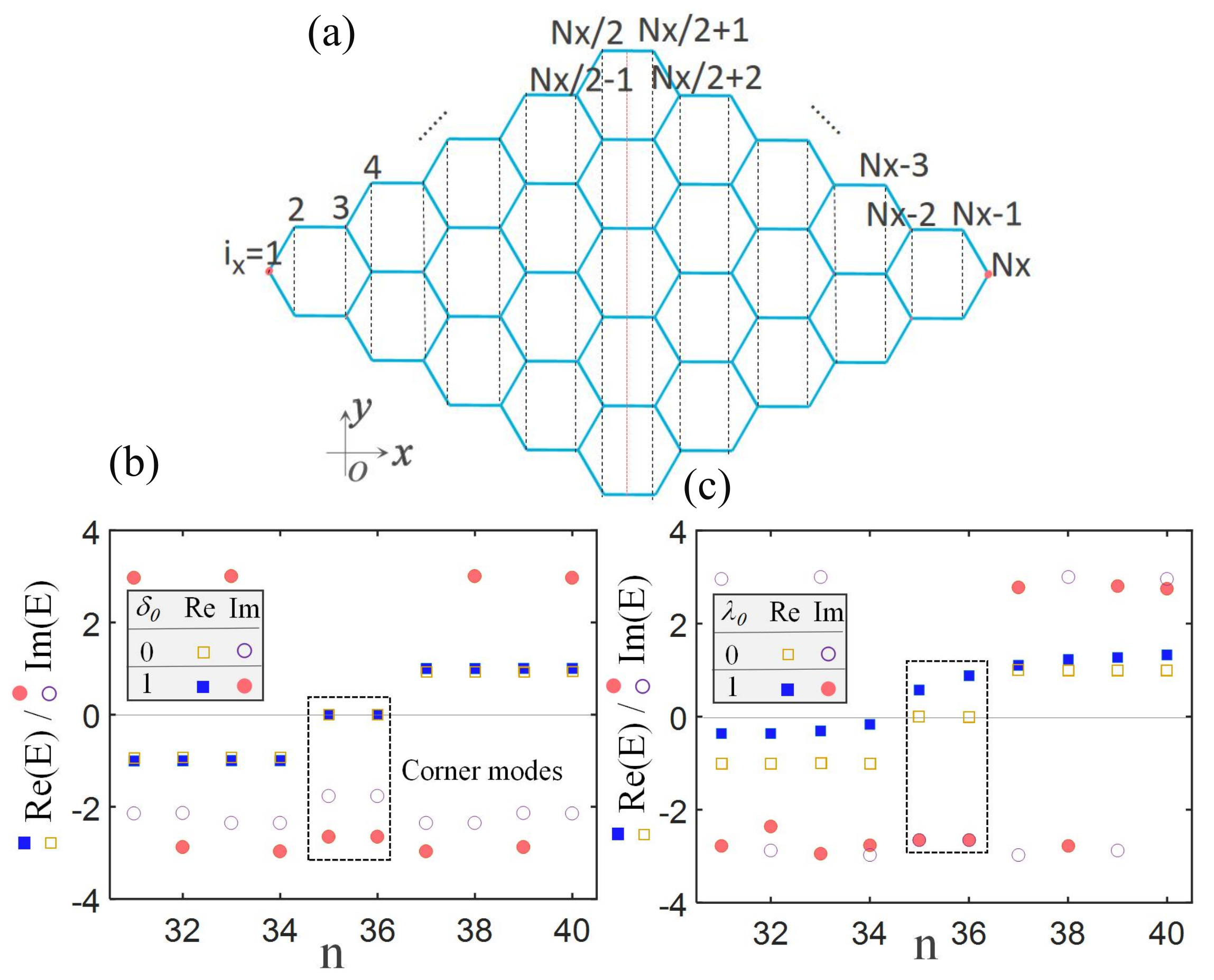}
\caption{(a) Illustration of honeycomb lattice with lattice index. The red
dashed line indicates a mirror. (b) Real and imaginary parts of
eigenenergies [$\func{Re}\left( E\right) $ and $\func{Im}\left( E\right) $]
vs the state index $n$ in the presence of random potentials (solid
squares and dots). The hollow squares and circles represents the case
without random potentials. The states in the dashed rectangle are photonic
corner modes with purely imaginary energies. The amplitude of random
potential $\protect\delta _{0}=1$. (c) Similar to (b), but with a random
potential that breaks reflection symmetry. The amplitude of random potential
$\protect\lambda _{0}=1$. Common parameters are $t=1$, $\protect\gamma =3$.}
\label{random}
\end{figure}
\label{app} We consider two cases in the following to show the protection
mechanism of higher-order topological states. First, we add general
perturbation terms with the hypothesized reflection symmetry as $%
H_{pa}=i\sum_{i}\delta _{i}a_{i}^{\dagger }a_{i}$, where the potential $%
\delta _{i}=\delta _{0}\kappa _{i_{x}}$, $\kappa _{i_{x}}=\kappa
_{N_{x}-i_{x}+1}\in \left[ 0,1\right] $ is a random number for $i_{x}\leq
N_{x}/2$, as shown in Fig. \ref{random}(a), and $\delta _{0}$ is the
amplitude of the random potential. Now the total Hamiltonian is $H_{\mathrm{T%
}}=H+H_{pa}$. Through numeric calculations, we find that localized corner
modes with purely imaginary energies still remain, where only an imaginary
energy shift appears for the topological corner states but the real
component of their energy is pinned at zero, as sketched in Fig. \ref{random}%
(b). Therefore, the topological corner states are robust against random
potentials (perturbations) with reflection symmetry.

Second, we add general random perturbation terms that break the
hypothesized reflection symmetry as $H_{pb}=\sum_{i}\lambda
_{i}a_{i}^{\dagger }a_{i}$, where $\lambda _{i}=\lambda _{0}\kappa _{i}$, $%
\kappa _{i}\in \left[ 0,1\right] $ is a random number. The numeric results
are shown in Fig. \ref{random}(c). We find that the corner states acquire a
non-zero real component of energy, and thus the corner states are no longer
topologically protected.

To sum up, the corner states are robust against perturbations with
reflection symmetry, but would disappear if the hypothesized reflection
symmetry is broken by the random potential.


\begin{thebibliography}{99}
\bibitem{Hasan2010} M. Z. Hasan and C. L. Kane, Colloquium: Topological
insulators, \href{https://doi.org/10.1103/RevModPhys.82.3045}{Rev. Mod.
Phys. \textbf{82}, 3045 (2010).}

\bibitem{Qi2011} X.-L. Qi and S.-C. Zhang, Topological insulators and
superconductors, \href{https://doi.org/10.1103/RevModPhys.83.1057}{Rev. Mod.
Phys. \textbf{83}, 1057 (2011).}

\bibitem{Ozawa2019} T. Ozawa, H. M. Price, A. Amo, N. Goldman, M. Hafezi, L.
Lu, M. C. Rechtsman, D. Schuster, J. Simon, O. Zilberberg, and I. Carusotto,
Topological photonics, \href{https://doi.org/10.1103/RevModPhys.91.015006}{%
Rev. Mod. Phys. \textbf{91}, 015006 (2019).}

\bibitem{Rudner2009} M. S. Rudner, and L. S. Levitov, Topological Transition
in a Non-Hermitian Quantum Walk, \href{https://doi.org/10.1103/PhysRevLett.103.248101}%
{Phys. Rev. Lett. \textbf{102}, 065703 (2009).}

\bibitem{Regensburger2013} A. Regensburger, M.-A. Miri, C. Bersch, J. N\"{a}%
ger, G. Onishchukov, D. N. Christodoulides, and U. Peschel, Observation of
Defect States in $\mathcal{PT}$-Symmetric Optical Lattices, \href{https://doi.org/10.1103/PhysRevLett.110.223902}%
{Phys. Rev. Lett. \textbf{110}, 223902 (2013).}


\bibitem{Lee2016} T. E. Lee, Anomalous Edge State in a Non-Hermitian
Lattice, \href{https://doi.org/10.1103/PhysRevLett.116.133903}{Phys. Rev.
Lett. \textbf{116}, 133903 (2016).}

\bibitem{Leykam2017} D. Leykam, K. Y. Bliokh C. Huang, Y. D. Chong, and F.
Nori, Edge Modes, Degeneracies, and Topological Numbers in Non-Hermitian
Systems, \href{https://doi.org/10.1103/PhysRevLett.118.040401}{Phys. Rev.
Lett. \textbf{118}, 040401 (2017).}

\bibitem{XuY2017} Y. Xu, S.-T. Wang, and L.-M. Duan, Weyl Exceptional Rings
in a Three-Dimensional Dissipative Cold Atomic Gas, \href{https://doi.org/10.1103/PhysRevLett.118.045701}%
{Phys. Rev. Lett. \textbf{118}, 045701 (2017).}

\bibitem{JinL2017} L. Jin, Topological phases and edge states in a
non-Hermitian trimerized optical lattice, \href{https://doi.org/10.1103/PhysRevA.96.032103}%
{Phys. Rev. A \textbf{96}, 032103 (2017).}

\bibitem{Shen2018} H. Shen, B. Zhen, and L. Fu, Topological Band Theory for
Non-Hermitian Hamiltonians,\href{https://doi.org/10.1103/PhysRevLett.120.146402}%
{ Phys. Rev. Lett. \textbf{120}, 146402 (2018).}

\bibitem{Flore2018} F. K. Kunst, E. Edvardsson, J. C. Budich, and E. J.
Bergholtz, Biorthogonal Bulk-Boundary Correspondence in Non-Hermitian
Systems,\href{https://doi.org/10.1103/PhysRevLett.121.026808}{ Phys. Rev.
Lett. \textbf{121}, 026808 (2018).}

\bibitem{Yao2018a} S. Yao and Z. Wang, Edge States and Topological
Invariants of Non-Hermitian Systems, \href{https://doi.org/10.1103/PhysRevLett.121.086803}%
{Phys. Rev. Lett. \textbf{121}, 086803 (2018).}

\bibitem{Yao2018b} S. Yao, F. Song, and Z. Wang, Non-Hermitian Chern Bands, Topological Phases of Non-Hermitian Systems,
\href{https://doi.org/10.1103/PhysRevLett.121.136802}{Phys. Rev. Lett.
\textbf{121}, 136802 (2018).}

\bibitem{Gong2018} Z. Gong, Y. Ashida, K. Kawabata, K. Takasan, S.
Higashikawa, and M. Ueda, \href{https://doi.org/10.1103/PhysRevX.8.031079}{%
Phys. Rev. X \textbf{8}, 031079 (2018).}

\bibitem{Yokomizo2019} K. Yokomizo and S. Murakami, Non-Bloch Band Theory of
Non-Hermitian Systems, \href{https://doi.org/10.1103/PhysRevLett.123.066404}{%
Phys. Rev. Lett. \textbf{123}, 066404 (2019).}

\bibitem{Turker2019} Z. O. Turker and C. Yuce, Open and closed boundaries in
non-Hermitian topological systems, \href{https://doi.org/10.1103/PhysRevA.99.022127}%
{Phys. Rev. A \textbf{99}, 022127 (2019).}

\bibitem{JinL2019} L. Jin and Z. Song, Bulk-boundary correspondence in a
non-Hermitian system in one dimension with chiral inversion symmetry, \href{https://doi.org/10.1103/PhysRevB.99.081103}%
{Phys. Rev. B \textbf{99}, 081103(R) (2019).}

\bibitem{LiuC2019} C.-H. Liu, H. Jiang, and S. Chen, Topological
classification of non-Hermitian systems with reflection symmetry, \href{https://doi.org/10.1103/PhysRevB.99.081103}%
{Phys. Rev. B \textbf{99}, 125103 (2019).}

\bibitem{Kawabata2019} K. Kawabata, S. Higashikawa, Z. Gong, Y. Ashida, and
M. Ueda, Topological unification of time-reversal and particle-hole
symmetries in non-Hermitian physics, \href{https://doi.org/10.1038/s41467-018-08254-y}%
{Nat. Commun. \textbf{10}, 297 (2019).}

\bibitem{Lieu2018} S. Lieu, Topological symmetry classes for non-Hermitian
models and connections to the bosonic Bogoliubov-de Gennes equation, \href{https://doi.org/10.1103/PhysRevB.98.115135}%
{Phys. Rev. B.\textbf{98}.115135 (2018).}

\bibitem{Poli2015} C. Poli, M. Bellec, U. Kuhl, F. Mortessagne, and H.
Schomerus, Selective enhancement of topologically induced interface states
in a dielectric resonator chain, \href{http://dx.doi.org/10.1038/ncomms7710}{%
Nat. Commun. \textbf{6}, 6710 (2015).}

\bibitem{Weimann2017} A. Weimann, M. Kremer, Y. Plotnik, Y. Lumer, S. Nolte,
K. Makris, M. Segev, M. Rechtsman, and A. Szameit, Topologically protected
bound states in photonic parity time-symmetric crystals, \href{http://dx.doi.org/10.1038/nmat4811}%
{Nat. Mater. \textbf{16}, 433 (2017).}

\bibitem{Feng2017} L. Feng, R. El-Ganainy, and L. Ge, Non-Hermitian
photonics based on parity-time symmetry, \href{https://doi.org/10.1038/s41566-017-0031-1}%
{Nat. Photonics \textbf{11}, 752 (2017).}

\bibitem{Xiao2017} L. Xiao, X. Zhan, Z. H. Bian, K. K. Wang, X. Zhang, X. P.
Wang, J. Li, K. Mochizuki, D. Kim, N. Kawakami, W. Yi, H. Obuse, B. C.
Sanders, and P. Xue, Observation of topological edge states in
parity-time-symmetric quantum walks, \href{http://dx.doi.org/10.1038/nphys4204}%
{Nat. Phys. \textbf{13}, 1117 (2017).}

\bibitem{Miri2019} M.-A. Miri, A. Al\`{u}, Exceptional points in optics and
photonics, \href{https://science.sciencemag.org/lookup/doi/10.1126/science.aar7709}%
{Science \textbf{363}, eaar7709 (2019).}

\bibitem{St-Jean2017} P. St-Jean, V. Goblot, E. Galopin, A. Lema\^{\i}tre,
T. Ozawa, L. L. Gratiet, I. Sagnes, J. Bloch, and A. Amo, Lasing in
topological edge states of a one-dimensional lattice,\href{http://dx.doi.org/10.1038/s41566-017-0006-2}%
{ Nat. Photonics \textbf{11}, 651 (2017).}

\bibitem{Zhao2018} H. Zhao, P. Miao, M. H. Teimourpour, S. Malzard, R.
El-Ganainy, H. Schomerus, and L. Feng, Topological hybrid silicon
microlasers, \href{http://dx.doi.org/10.1038/s41467-018-03434-2}{Nat.
Commun. \textbf{9}, 981 (2018).}

%

\bibitem{Efremidis2002} N. K. Efremidis, S. Sears, D. N. Christodoulides, J.
W. Fleischer, and M. Segev, Discrete solitons in photorefractive optically
induced photonic lattices, \href{https://doi.org/10.1103/PhysRevE.66.046602}{%
Phys. Rev. E \textbf{66}, 046602 (2002).}

\bibitem{Bartal2005} G. Bartal, O. Cohen, H. Buljan, J. W. Fleischer, O.
Manela, and M. Segev, Brillouin Zone Spectroscopy of Nonlinear Photonic
Lattices, \href{https://doi.org/10.1103/PhysRevLett.94.163902}{Phys. Rev.
Lett. \textbf{94}, 163902 (2005).}

\bibitem{Peleg2007} O. Peleg, G. Bartal, B. Freedman, O. Manela, M. Segev,
and D. N. Christodoulides, Conical Diffraction and Gap Solitons in Honeycomb
Photonic Lattices, \href{https://doi.org/10.1103/PhysRevLett.98.103901}{%
Phys. Rev. Lett. \textbf{98}, 103901(2007).}

\bibitem{Sepkhanov2007} R. A. Sepkhanov, Y. B. Bazaliy, and C. W. J.
Beenakker, Extremal transmission at the Dirac point of a photonic band
structure, \href{https://doi.org/10.1103/PhysRevA.75.063813}{Phys. Rev. A
\textbf{75}, 063813 (2007).}

\bibitem{Treidel2008} O. Bahat-Treidel, O. Peleg, and M. Segev, Symmetry
breaking in honeycomb photonic lattices, \href{https://doi.org/10.1364/OL.33.002251}%
{Opt. Lett. \textbf{33}, 2251 (2008).}

\bibitem{Treidel2010} O. Bahat-Treidel, O. Peleg, M. Grobman, N. Shapira, M.
Segev, and T. Pereg-Barnea, Klein Tunneling in Deformed Honeycomb Lattices,
\href{http://dx.doi.org/10.1103/PhysRevLett.104.063901}{Phys. Rev. Lett.
\textbf{104}, 063901 (2010).}

\bibitem{Polini2013} M. Polini, F. Guinea, M. Lewenstein, H. C. Manoharan,
and V. Pellegrini, Artificial honeycomb lattices for electrons, atoms and
photons, \href{https://doi.org/10.1038/nnano.2013.161}{Nat. Nanotech.
\textbf{8}, 625 (2013)}

\bibitem{Plotnik2014} Y. Plotnik, M. C. Rechtsman, D. Song, M. Heinrich, J.
M. Zeuner, S. Nolte, Y. Lumer, N. Malkova, J. Xu, A. Szameit, Z. Chen and M.
Segev, Observation of unconventional edge states in `photonic graphene',
\href{https://doi.org/10.1038/nmat3783}{Nat. Mater. \textbf{13}, 57 (2014).}
\

\bibitem{Oztas2018} Z. Oztas and C. Yuce, Spontaneously broken particle-hole
symmetry in photonic graphene with gain and loss, \href{https://doi.org/10.1103/PhysRevA.98.042104}%
{Phys. Rev. A \textbf{98,} 042104 (2018).}

\bibitem{Kobayashi2005} Y. Kobayashi, K.-i. Fukui, T. Enoki, K. Kusakabe,
and Y. Kaburagi, Observation of zigzag and armchair edges of graphite using
scanning tunneling microscopy and spectroscopy, \href{https://doi.org/10.1103/PhysRevB.71.193406}%
{Phys. Rev. B \textbf{71}, 193406 (2005)}

\bibitem{Rechtsman2013} M. C. Rechtsman, Y. Plotnik, J. M. Zeuner, D. Song,
Z. Chen, A. Szameit, and M. Segev, Topological Creation and Destruction of
Edge States in Photonic Graphene, \href{https://doi.org/10.1103/PhysRevLett.111.103901}%
{Phys. Rev. Lett. \textbf{111}, 103901 (2013).}

\bibitem{Benalcazar2017} W. A. Benalcazar, B. A. Bernevig, and T. L. Hughes,
Quantized electric multipole insulators, \href{http://dx.doi.org/10.1126/science.aah6442}%
{Science \textbf{357}, 61(2017).}

\bibitem{Langbehn2017} J. Langbehn, Y. Peng, L. Trifunovic, F. von Oppen,
and P. W. Brouwer, Refection-Symmetric Second-Order Topological Insulators
and Superconductors, \href{http://dx.doi.org/10.1103/PhysRevLett.119.246401}{%
Phys. Rev. lett. \textbf{119}, 246401 (2017).}

\bibitem{Song2017} Z. Song, Z. Fang, and C. Fang, ($d-2$)-Dimensional Edge
States of Rotation Symmetry Protected Topological States, \href{http://dx.doi.org/10.1103/PhysRevLett.119.246402}%
{Phys. Rev. lett. \textbf{119}, 246402 (2017).}

\bibitem{Ezawa2018} M. Ezawa, Minimal models for Wannier-type higher-order
topological insulators and phosphorene, \href{https://doi.org/10.1103/PhysRevB.98.045125}%
{Phys. Rev. B \textbf{98}, 045125 (2018).}

\bibitem{Noh2018} J. Noh, W. A. Benalcazar, S. Huang, M. J. Collins, K. P.
Chen, T. L. Hughes, and M. C. Rechtsman, Topological protection of photonic
mid-gap defect modes,\ \href{https://doi.org/10.1038/s41566-018-0179-3}{Nat.
Photonics \textbf{12}, 408(2018).}

\bibitem{ChenX2019} X.-D. Chen, W.-M. Deng, F.-L. Shi, F.-L. Zhao, M. Chen,
and J.-W. Dong, Direct Observation of Corner States in Second-Order
Topological Photonic Crystal Slabs, \href{http://dx.doi.org/10.1103/PhysRevLett.119.246402}%
{Phys. Rev. Lett. \textbf{122}, 233902 (2019).}

\bibitem{Hassan2019} A. E. Hassan, F. K. Kunst, A. Moritz, G. Andler, E. J.
Bergholtz, and M. Bourennane, Corner states of light in photonic waveguides,
\href{https://doi.org/10.1038/s41566-019-0519-y}{Nat. Photonics \textbf{13},
697 (2019).}

\bibitem{Xie2018} B.-Y. Xie, H.-F. Wang, H.-X. Wang, X.-Y. Zhu, J.-H. Jiang,
M.-H. Lu, and Y.-F. Chen, Second-order photonic topological insulator with
corner states, \href{https://doi.org/10.1103/PhysRevB.98.205147}{Phys. Rev.
B \textbf{98}, 205147 (2018).}

\bibitem{liut2019} T. Liu, Y.-R. Zhang, Q. Ai, Z. Gong, K. Kawabata, M.
Ueda, and F. Nori, Second-Order Topological Phases in Non-Hermitian Systems,
\href{https://doi.org/10.1103/PhysRevLett.122.076801}{Phys. Rev. Lett.
\textbf{122}, 076801(2019).}

\bibitem{Lee2019} C. H. Lee, L. Li, and J. Gong, Hybrid Higher-Order
Skin-Topological Modes in Nonreciprocal Systems, \href{https://doi.org/10.1103/PhysRevLett.123.016805}%
{Phys. Rev. Lett. \textbf{123}, 016805 (2019).}

\bibitem{Luo2019} X.-W. Luo and C. Zhang, Higher-Order Topological Corner
States Induced by Gain and Loss, \href{https://doi.org/10.1103/PhysRevLett.123.073601}%
{Phys. Rev. Lett. \textbf{123}, 073601 (2019).}

\bibitem{Benalcazar2019} W. A. Benalcazar, T. Li, and T. L. Hughes,
Quantization of fractional corner charge in $C_n$-symmetric higher-order
topological crystalline insulators, \href{https://doi.org/10.1103/PhysRevB.99.245151}%
{Phys. Rev. B \textbf{99},245151 (2019).}

\bibitem{Hou2019} J. Hou, Y.-J. Wu, and C. Zhang, Non-Hermitian topological
phase transitions for quantum spin Hall insulators, \href{https://arxiv.org/abs/1910.14606}%
{arXiv: 1910.14606 (2019).}


\bibitem{Harari2018} G. Harari, M. A. Bandres, Y. Lumer, M. C. Rechtsman, Y.
D. Chong, M. Khajavikhan, D. N. Christodoulides, and M. Segev, Topological
insulator laser: Theory, \href{https://science.sciencemag.org/content/359/6381/eaar4003}%
{Science \textbf{359}, eaar4003 (2018)}

\bibitem{Bandres2018} M. A. Bandres, S. Wittek, G. Harari, M. Parto, J. Ren,
M. Segev, D. N. Christodoulides, M. Khajavikhan, Topological insulator
laser: Experiments, \href{https://science.sciencemag.org/content/359/6381/eaar4005#aff-1}%
{Science \textbf{359}, eaar4005 (2018)}

\bibitem{Parto2018} M. Parto, S. Wittek, H. Hodaei, G. Harari, M. A.
Bandres, J. Ren, M. C. Rechtsman, M. Segev, D. N. Christodoulides, and M.
Khajavikhan, Edge-Mode Lasing in 1D Topological Active Arrays, \href{https://doi.org/10.1103/PhysRevLett.123.066404}%
{Phys. Rev. Lett. \textbf{120}, 113901 (2018).}

\bibitem{Bahari2017} B. Bahari, A. Ndao, F. Vallini, A. E. Amili, Y.
Fainman, and B. Kant\'{e}, Nonreciprocal lasing in topological cavities of
arbitrary geometries, \href{https://science.sciencemag.org/content/358/6363/636}%
{Science \textbf{358}, 636 (2017).}

\bibitem{Kunst2019} F. K. Kunst, and V. Dwivedi, Non-Hermitian systems and
topology: A transfer-matrix perspective, \href{https://doi.org/10.1103/PhysRevB.99.245116}%
{Phys. Rev. B \textbf{99}, 245116 (2019).}

\bibitem{Zeuner2015} J. M. Zeuner, M. C. Rechtsman, Y. Plotnik, Y. Lumer, S.
Nolte, M. S. Rudner, Observation of a Topological Transition in the Bulk of
a Non-Hermitian System, \href{http://dx.doi.org/10.1103/PhysRevLett.115.040402}%
{Phys. Rev. Lett. \textbf{115}, 040402 (2015).}

\bibitem{Yucea2018} C. Yuce, Edge states at the interface of non-Hermitian
systems, \href{https://doi.org/10.1103/PhysRevA.97.042118}{Phys. Rev. A
\textbf{97}, 042118 (2018).}

\bibitem{Okugawa2019} R. Okugawa and T. Yokoyama, Topological exceptional
surfaces in non-Hermitian systems with parity-time and parity-particle-hole
symmetries, \href{https://doi.org/10.1103/PhysRevB.99.041202}{Phys. Rev. B
\textbf{99}, 041202 (2019).}

\bibitem{Kawabata2018} K. Kawabata, Y. Ashida, H. Katsura, and M. Ueda,
Parity-time-symmetric topological superconductor, \href{https://doi.org/10.1103/PhysRevB.98.085116}%
{Phys. Rev. B \textbf{98}, 085116 (2018).}
\end{thebibliography}
\end{document}